\documentclass[twocolumn,prb,preprintnumbers,amsmath,amssymb,showpacs,floatfix]{revtex4}

\usepackage{graphicx}
\usepackage{subfig}
\usepackage{dcolumn}
\usepackage{bm}
\begin{document}
\title{Symmetry analysis of magneto-optical effects: The case of x-ray diffraction and x-ray absorption at the transition metal $L_{2,3}$ edge}

\author{M. W. Haverkort}
  \affiliation{Max Planck Institute for Solid State Research, Heisenbergstra{\ss}e 1, D-70569 Stuttgart, Germany}
\author{N. Hollmann}
  \affiliation{ II. Physikalisches Institut, University of Cologne, Z{\"u}lpicher Stra{\ss}e 77, D-50937 K{\"o}ln, Germany}
\author{I. P. Krug}
  \affiliation{ Institut f{\"u}r Festk{\"o}rperforschung IFF-9, Forschungszentrum J{\"u}lich GmbH, D-52425 J{\"u}lich, Germany}
\author{A. Tanaka}
  \affiliation{Department of Quantum Matter, ADSM, Hiroshima University - Higashi-Hiroshima 739-8530, Japan}

\date{\today}

\begin{abstract}

A general symmetry analysis of the optical conductivity or scattering tensor is used to rewrite the conductivity tensor as a sum of fundamental spectra multiplied by simple functions depending on the local magnetization direction. Using this formalism, we present several numerical examples at the transition metal $L_{2,3}$ edge. From these numerical calculations we can conclude that large deviations from the magneto-optical effects in spherical symmetry are found. These findings are in particular important for resonant x-ray diffraction experiments where the polarization dependence and azimuthal dependence of the scattered Bragg intensity is used to determine the local ordered magnetization direction.
\end{abstract}


 \maketitle

Resonant x-ray diffraction or reflectivity (RXD) has developed into a powerful method to study charge, orbital and magnetic ordering in transition metal compounds and artificially created superlattices. In principle the resonant energy dependence of the scattering profile and its polarization and azimuthal dependence contain the information about the local magnetic, charge, and orbital order. Obtaining this information from such spectra is often still a theoretical and experimental challenge.

One of the more straightforward methods to analyze magnetic Bragg reflections has been presented by Hannon \textit{et al.}\cite{Hannon88} and several other authors. \cite{Blume85, Platzman70, DeBergevin81, Blume88, Lovesey87, Rennert93, Hill95} They showed that the azimuthal intensity dependence can be related to the local orientation of the magnetic moment by a simple relation. The scattered intensity is proportional to $\left|(\bm{\varepsilon}_{in}\times\bm{\varepsilon}_{out}^*)\cdot\bm{\hat{m}}\right|^2$, whereby $\bm{\varepsilon}_{in (out)}$ is the polarization of the incoming (outgoing) light and $\bm{\hat{m}}$ is a unit vector in the direction of the magnetization. These relations are used relatively often in the analyses of RXD nowadays. \cite{HannonUse} The crux is that these relations are derived in spherical symmetry and strictly speaking hold only in spherical symmetry. An extended formalism has been presented, \cite{Carra94, Matteo03} but not seen a large audience so far. The question the present paper tries to answer is: how large are the changes of the scattered intensity when the effects of the real crystal symmetry is included and when is it important to include the real crystal symmetry. In other words we answer the question when one can use the formulas derived in spherical symmetry and when the true crystal symmetry has to be included. We will focus on purely magnetic Bragg reflections and neglect spin-orbit coupling for the initial or ground-state when doing the numerical calculations.

A hint for the importance of including the correct crystal symmetry might be obtained from absorption spectroscopy. By the optical theorem \cite{Newton76} it is well known that the scattering tensor $F$  and the conductivity tensor $\sigma$ are related by a factor of $\omega$ ($F\propto\omega\sigma$). Therefore all magneto-optical effects known from absorption spectroscopy should return in diffraction or reflectivity experiments. Recently it has been realized that the symmetric part of the conductivity tensor due to magnetism i.e. the magnetic linear dichroism effect is not given by a single spectrum as in spherical symmetry, but behaves more complicated in cubic or lower symmetries\cite{Arenholz06, Arenholz07, Laan08, Kunes03}. For an antiferromagnetic Bragg reflection the symmetric part of the scattering tensor does not contribute to the diffracted intensity. How the antisymmetric part of the conductivity tensor behaves for less than spherical symmetry is not well known in the x-ray regime. For infra-red and visible wave-lengths there are several experiments and numerical calculations showing that for systems with less than cubic symmetry, the magnetic part of the optical conductivity tensor $\sigma$ becomes considerably different from what is expected in spherical symmetry. \cite{Oppeneer95, Uspenskii96, Uba98} One might thus expect that also for x-ray wavelengths the spectral line-shape of circular dichroism and magnetic Bragg reflections should be crucially influenced by the crystal symmetry.

In the present paper, we show several calculations at the transition metal $L_{2,3}$ edge in order to exemplify when deviations from spherical symmetry become important. To present the results in a consistent picture, we first introduce a symmetry analysis of the magnetization direction dependence of the conductivity tensor or scattering tensor. This allows us to define a few fundamental spectra that describe the entire optical response of a system independent of the magnetization direction. We will assume a certain symmetry in the paramagnetic phase and in the ordered phase an additional, symmetry breaking sub-lattice magnetization in an arbitrary direction, given by $\bm{m}\equiv[m_x,m_y,m_z]$. \cite{vector} The tensor formalism used is in principle valid for any type of magneto-optical experiment and closely related, though not entirely equivalent, to previous publications. \cite{Hannon88, Blume85, Platzman70, DeBergevin81, Blume88, Lovesey87, Carra94, Matteo03,Arenholz06, Arenholz07, Laan08, Kunes03,Oppeneer95, Uspenskii96, Uba98} The formalism can be used for the description of the crystal orientation dependence of the Kerr angle in the infra-red regime on CrO$_{2}$ \cite{Uspenskii96} or the magnetic Bragg diffraction of the cuprates at 930-950 eV.

The first section presents a general derivation of the scattering tensor or conductivity tensor as a function of magnetization direction for several different point-group symmetries. In the second section we will show how these tensors can be rewritten to simple dot products relating the scattered intensity to the polarization of the incoming and outgoing light in the same way as previously presented\cite{Hannon88} in spherical symmetry. The third section is reserved for calculations of the fundamental spectra of several example materials in cubic point-group symmetry (the symmetry relates to the paramagnetic phase). The fourth section presents results in tetragonal symmetry. Within the conclusions we will discuss when deviation from spherical symmetry becomes visible in experiment and how the magnetization direction can be deduced from resonant x-ray diffraction experiments.

\section{Scattering tensor of systems with arbitrary magnetization direction}

The general scattering tensor of a system of triclinic symmetry and an arbitrary magnetization is given by
\begin{equation}
F(\omega)=\left(
  \begin{array}{ccc}
    F_{xx}(\omega) & F_{yx}(\omega) & F_{zx}(\omega) \\
    F_{xy}(\omega) & F_{yy}(\omega) & F_{zy}(\omega) \\
    F_{xz}(\omega) & F_{yz}(\omega) & F_{zz}(\omega) \\
  \end{array}
\right)
\end{equation}
which defines a non-symmetric, non-Hermitian, complex tensor. Naturally this tensor can be diagonalized to give a complex basis defining the principal axes and a diagonal tensor. In general the principal axes will be $\omega$ dependent. Even in high symmetries there will already be a strong $\omega$ dependence on the principal axes when a magnetization in an arbitrary direction is introduced. It is therefore often more intuitive to stay in a Cartesian basis for the scattering tensor, with $x$, $y$, and $z$ aligned along high symmetry crystal axes. 

Based on general symmetry arguments the scattering tensor simplifies. Let $\Gamma$ be the set of symmetry operations present in the paramagnetic phase of a material. In the magnetically ordered phase the local moments order in the $[m_x,m_y,m_z]$ direction. We will assume that the crystal structure does not change between the paramagnetic and ordered phase. In the case that the local moment direction does not coincide with one of the high symmetry axes of the paramagnetic phase, only the identity operation is left of the original symmetry operations as all other symmetry operations will rotate the local moment. One can however define a new set of symmetry operations $\Gamma'$ that rotates first the entire system and then the local moment, i.e. the spin and orbital momentum, back. This would be the same as rotating the system, but not the local moment.

In order to define symmetry operations that rotate the system but not the local moment, the scattering tensor needs to be written as a product of a function that is independent of the direction of the local moment, but depends on $\omega$ and the polarization and a function that depends purely on the direction of the local moment. This can be done by expanding each element of the scattering tensor on spherical harmonics in the coordinates of the local moment.
\begin{equation}
F(\theta,\phi)=\sum_{k=0}^{\infty}\sum_{m=-k}^{k}\left(
  \begin{array}{ccc}
    F_{xx}^{km} & F_{yx}^{km} & F_{zx}^{km} \\
    F_{xy}^{km} & F_{yy}^{km} & F_{zy}^{km} \\
    F_{xz}^{km} & F_{yz}^{km} & F_{zz}^{km} \\
  \end{array}
\right)Y_{km}(\theta,\phi)
\end{equation}
$\theta$ and $\phi$ define the direction of the local moment ($m_x=|\bm{m}|\cos(\phi)\sin(\theta)$, $m_y=|\bm{m}|\sin(\phi)\sin(\theta)$, and $m_z=|\bm{m}|\cos(\theta)$), $Y_{k,m}$ is a spherical harmonic function and $F_{i,j}$ is the $(i,j)$ component of the scattering tensor on a basis of linear polarized light in the coordinate system of the crystal ($\varepsilon_i,\varepsilon_j\in \{\varepsilon_x,\varepsilon_y,\varepsilon_z\}$). Note that both the local magnetization direction as well as the polarization are expressed in the same Cartesian coordinate system.

Finding the symmetry allowed components of the scattering tensor implies going through all symmetry operations possible and solving the equation $\Gamma' F=F$. This will lead to a set of allowed values for the expansion coefficients. For example a ${C'}_4^z$ operation acting on $F(\theta,\phi)$ i.e. rotating the system, but not rotating the magnetization direction, would give
\begin{eqnarray}
{C'}_4^z F(\theta,\phi)=&\\\nonumber \sum_{k=0}^{\infty}\sum_{m=-k}^{k}&\left(
  \begin{array}{ccc}
    F_{yy}^{km} &-F_{xy}^{km} & F_{zy}^{km} \\
   -F_{yx}^{km} & F_{xx}^{km} &-F_{zx}^{km} \\
    F_{yz}^{km} &-F_{xz}^{km} & F_{zz}^{km} \\
  \end{array}
\right)Y_{km}(\theta,\phi-\frac{1}{2}\pi)
\end{eqnarray}
Which leads to sets of equations of the form $\sum_{k=0}^{\infty}\sum_{m=-k}^{k} F_{yx}^{km} Y_{km}(\theta,\phi) = \sum_{k=0}^{\infty}\sum_{m=-k}^{k}-F_{xy}^{km} Y_{km}(\theta,\phi-1/2\pi)$, which have to be solved.

Let us first discuss spherical symmetry. In spherical symmetry there should be no $\phi$ dependence and only terms with $m=0$ remain. Furthermore the infinite sum truncates at $k=2$ due to the triangular equations. This leaves only the expansion coefficients proportional to $Y_{00}$, $Y_{10}$ and $Y_{20}$. For a local moment in the $z$ direction ($\bm{\hat{m}}=[001]$) one gets the known result:
\begin{equation}
F_{[001]}=\left(
  \begin{array}{ccc}
    F^{(0)}-\frac{1}{3}F^{(2)} & F^{(1)}                & 0                  \\
   -F^{(1)}                & F^{(0)}-\frac{1}{3}F^{(2)} & 0                  \\
    0                  & 0                  & F^{(0)}+\frac{2}{3}F^{(2)} \\
  \end{array}
\right)
\end{equation}
and for a local moment in the $(x\equiv m_x/|\bm{m}|,y\equiv m_y/|\bm{m}|,z\equiv m_z/|\bm{m}|)$ direction:
\begin{widetext}
\begin{equation}
F_{[xyz]}=\left(
  \begin{array}{ccc}
    F^{(0)}+(x^2-\frac{1}{3})F^{(2)} & (z)F^{(1)}+(x y)F^{(2)}          &-(y)F^{(1)}+(x z)F^{(2)}      \\
   -(z)F^{(1)}+(x y)F^{(2)}          & F^{(0)}+(y^2-\frac{1}{3})F^{(2)} & (x)F^{(1)}+(y z)F^{(2)}      \\
    (y)F^{(1)}+(x z)F^{(2)}          &-(x)F^{(1)}+(y z)F^{(2)}          & F^{(0)}+(z^2-\frac{1}{3})F^{(2)} \\
  \end{array}
\right)
\end{equation}
\end{widetext}
The result for a local moment in an arbitrary direction could also be obtained from the tensor of the local moment in the $z$ direction by rotating the scattering tensor.
\begin{equation}
R=\left(
  \begin{array}{ccc}
    \cos(\phi)  &-\sin(\phi)  & 0   \\
    \sin(\phi)  & \cos(\phi)  & 0   \\
    0           & 0           & 1   \\
  \end{array}
\right)\cdot\left(
  \begin{array}{ccc}
    \cos(\theta) & 0 & \sin(\theta) \\
    0            & 1 & 0   \\
   -\sin(\theta) & 0 & \cos(\theta) \\
  \end{array}
\right)
\end{equation}
then
\begin{equation}
F_{[xyz]}=R F_{[001]} R^{T}
\end{equation}
$F^{(1)}$ is related to the gyromagnetic vector by $F^{(1)}/\omega\propto\imath\bm{g}$ and describes the circular dichroism or the Faraday effect. The $F^{(2)}$ components describe the linear dichroism. The Magneto Optical Kerr Effect is given by both the $F^{(1)}$ and $F^{(2)}$ spectra depending on the experimental geometry.

In symmetries lower than spherical the expansion of the spin direction in spherical harmonics does not truncate at finite $k$. This has often been neglected previously, but is quite obvious as angular momentum is not a conserved quantum number in real crystals. Elements like $F^{(3)}$, $F^{(4)}$, etc. are allowed by symmetry. There is thus in principle an infinite number of fundamental spectra. Not all of them are important and most of them will be very small. Below we will discus the higher order expansions in more detail and give several numerical examples for realistic parameters. Furthermore, as previously discussed by Carra and Thole \cite{Carra94} the fundamental spectra of order $k$ will branch according to their symmetry representations in the corresponding point-group.

In cubic symmetry ($O_h$) the scattering tensor becomes ($C_4 \parallel <001>$):
\begin{widetext}
\begin{eqnarray}
&&F_{[xyz]}=\\ \nonumber&& \left(
  \begin{array}{ccc}
    F^{(0)}_{a_{1g}}+(x^2\!-\!\frac{1}{3})F^{(2)}_{e_g} & z F^{(1)}_{t_{1u}}+z(z^2\!-\!\frac{3}{5}) F^{(3)}_{t_{1u}}+x y F^{(2)}_{t_{2g}}         &-y F^{(1)}_{t_{1u}}-y(y^2\!-\!\frac{3}{5}) F^{(3)}_{t_{1u}}+x z F^{(2)}_{t_{2g}}         \\
   -z F^{(1)}_{t_{1u}}-z(z^2\!-\!\frac{3}{5}) F^{(3)}_{t_{1u}}+x y F^{(2)}_{t_{2g}}         & F^{(0)}_{a_{1g}}+(y^2\!-\!\frac{1}{3})F^{(2)}_{e_g} & x F^{(1)}_{t_{1u}}+x(x^2\!-\!\frac{3}{5})F^{(3)}_{t_{1u}}+y z F^{(2)} _{t_{2g}}        \\
     y F^{(1)}_{t_{1u}}+ y(y^2\!-\!\frac{3}{5})F^{(3)}_{t_{1u}}+ x z F^{(2)}_{t_{2g}}         &- x F^{(1)}_{t_{1u}}- x(x^2\!-\!\frac{3}{5}) F^{(3)}_{t_{1u}}+ y z F^{(2)}_{t_{2g}}         & F^{(0)}_{a_{1g}}+(z^2-\frac{1}{3})F^{(2)}_{e_g} \\
  \end{array}
\right)
\end{eqnarray}
\end{widetext}
Whereby the expansion series can be continued by summing for the diagonals the $a_{1g}$ and $e_g$ cubic harmonics of order $k$ multiplied by a fundamental spectrum and for the off diagonal components the $t_{1u}$ and $t_{2g}$ cubic harmonics.

The important change between cubic and spherical symmetry is that $F^{(2)}$ becomes different ($F^{(2)}_{t_{2g}}$ or $F^{(2)}_{e_g}$) for diagonal and off diagonal elements in $F$. $F^{(2)}_{e_g}$ defines the magnetic linear dichroic spectrum one measures if the sample is magnetized along a $C_{4}$ direction, whereas $F^{(2)}_{t_{2g}}$ defines the magnetic linear dichroic spectrum one measures if the sample is magnetized along a $C_{3}$ direction. For very small deviations from spherical symmetry $F^{(2)}_{t_{2g}}$ must be roughly equal to $F^{(2)}_{e_g}$, however as we will show below by several numerical examples one finds for real systems that $F^{(2)}_{e_g}$ and $F^{(2)}_{t_{2g}}$ are very different. On top of that one finds that for realistic parameters and large local moments the contribution of $F^{(3)}_{t_{1u}}$ can not be neglected.

In tetragonal symmetry ($D_{4h}$) the scattering tensor becomes ($C_4 \parallel [001], C_2 \parallel <100>$).
\begin{widetext}
\begin{eqnarray}
&&F_{[xyz]}=\\ \nonumber&&\left(
  \begin{array}{ccc}
    F^{(0)}_{a_{1g}^B}+\frac{1}{2}(x^2\!-\!y^2)F^{(2)}_{b_{1g}}-\frac{1}{2}(z^2\!-\!\frac{1}{3})F^{(2)}_{a_{1g}^B} & z F^{(1)}_{a_{2u}}+z(z^2\!-\!\frac{3}{5}) F^{(3)}_{a_{2u}}+x y F^{(2)}_{b_{2g}} &-yF^{(1)}_{e_u}-y(y^2\!-\!\frac{3}{5}) F^{(3)}_{e_u}+x z F^{(2)}_{e_g} \\
   -z F^{(1)}_{a_{2u}}-z(z^2\!-\!\frac{3}{5}) F^{(3)}_{a_{2u}}+ x y F^{(2)}_{b_{2g}} & F^{(0)}_{a_{1g}^B}-\frac{1}{2}(x^2\!-\!y^2)F^{(2)}_{b_{1g}}-\frac{1}{2}(z^2\!-\!\frac{1}{3})F^{(2)}_{a_{1g}^B} & x F^{(1)}_{e_u}+x(x^2\!-\!\frac{3}{5}) F^{(3)}_{e_u}+ y z F^{(2)}_{e_g} \\
     y F^{(1)}_{e_u}+y(y^2\!-\!\frac{3}{5}) F^{(3)}_{e_u}+ x z F^{(2)}_{e_g}   &- x F^{(1)}_{e_u}-x(x^2\!-\!\frac{3}{5}) F^{(3)}_{e_u}+ y z F^{(2)}_{e_g}                                                      & F^{(0)}_{a_{1g}^A}+(z^2\!-\!\frac{1}{3})F^{(2)}_{a_{1g}^A} \\
  \end{array}
\right)
\end{eqnarray}
\end{widetext}
The difference between $F^{(0)}_{a_{1g}^A}$ and $F^{(0)}_{a_{1g}^B}$ defines the natural linear dichroic spectra, also present in a paramagnetic sample.

There are five fundamental spectra up to order $k=2$ that describe the magnetic linear dichroism. Let us give examples how to obtain each one of these spectra. Placing the polarization into the $z$ direction, the dichroism resulting from changing the magnetization direction from $x$ to $z$ is described by $F_{a_{1g}^A}^{(2)}$. When the polarization is along $[110]$ and the magnetisation changes from $x$ to $z$, $F_{a_{1g}^B}^{(2)}$ describes the magnetic linear dichroism. Placing the polarization along $x$, changing the magnetization from $x$ to $y$ will result in a magnetic linear dichroism determined by $F_{b_{1g}}^{(2)}$. There are two off-diagonal elements ($F_{b_{2g}}^{(2)}$ and $F_{e_g}^{(2)}$) that define the dichroism when the magnetization is in the $[111]$ direction: $F_{b_{2g}}^{(2)}$ for the dichroism between polarizations along $[110]$ and $[1\overline{1}0]$, $F_{e_{g}}^{(2)}$ for polarizations along $[101]$ and $[10\overline{1}]$.

The isotropic spectrum, i.e. the spectrum measured on a powdered sample, is given by the trace of the conductivity tensor. The isotropic spectrum on a single crystal is measured by averaging three orthogonal polarization directions. For magnetic systems in cubic symmetry the trace is independent of the magnetization direction. The isotropic spectrum of a para-magnet is thus equal to the isotropic spectrum of the magnetically ordered system. It is interesting to note that the trace of the conductivity tensor in $D_{4h}$ symmetry becomes dependent on the magnetization direction. This is due to the $F^{(2)}_{e_g}$ spectrum in cubic symmetry that branches to an $F^{(2)}_{a_{1g}^A}$ spectrum for $z$ polarization and to an $F^{(2)}_{a_{1g}^B}$ and $F^{(2)}_{b_{1g}}$ spectrum for $x$ or $y$ polarization. This finding does not contradict that the spectrum of a paramagnet only depends on $k=0$ components. It is also valid that the spectrum of a multidomain sample averaged over all possible spin directions such that the final magnetic symmetry is at least cubic equals the spectrum of a para-magnet. The dependence of the trace of the conductivity tensor on the spin direction might therefore come as a surprise. However it is allowed by symmetry, which can be understood with an example. For a $S=1$ system one has three low energy eigen-states, $S_z=1$, $0$ or $-1$. With these three eigen-states any spin direction can be created. Within cubic symmetry these three states still belong to the same irreducible representation ($t_{1u}$) and rotating the spin therefore does not change the symmetry of the ground-state. In tetragonal symmetry the $S_z=\pm1$ states however have a different symmetry ($e_u$) from the $S_z=0$ state ($a_{2u}$). Naively one might expect that this is unimportant as these states will be degenerate as long as spin-orbit coupling for the $d$ shell is neglected in the initial state. When spin-orbit coupling is included one might still expect that the splitting will be smaller than the magnetic field applied or the internal exchange fields present in the sample. But formally the ground-state symmetry changes depending on the magnetization direction and therefore the isotropic spectra (the trace of the conductivity tensor) might have a different line-shape depending on having an $e_g$ ($S_z=\pm1$) or $a_{2u}$ ($S_z=0$) ground-state. The size of these effects will be shown below by numerical calculations.

In tetragonal symmetry the antisymmetric part of order $k=1$ can not be represented by a single spectrum. The circular dichroic spectrum will be different for a magnetization direction parallel or perpendicular to the $C_4$ axes. In scattering experiments this fundamental spectrum determines the line-shape of the first order Bragg reflection of an antiferromagnetically ordered crystal. It will therefore have a large impact on the interpretation of scattering data.

In orthorhombic symmetry ($D_{2h}$) the scattering tensor becomes ($C_2 \parallel <001>$):
\begin{widetext}
\begin{eqnarray}
&&F_{[xyz]}=\\ \nonumber&&\left(
  \begin{array}{ccc}
    F^{(0)}_{a_g^{xx}}+\!(x^2\!-\!\frac{1}{3})F^{(2)}_{a_g^{xxA}}+\!(y^2\!-\!z^2)F^{(2)}_{a_g^{xxB}} & z F^{(1)}_{b_{1u}}+z(z^2\!-\!\frac{3}{5}) F^{(3)}_{b_{1u}}+x y F^{(2)}_{b_{1g}}    &-y F^{(1)}_{b_{2u}}-y(y^2\!-\!\frac{3}{5}) F^{(3)}_{b_{2u}}+x z F^{(2)}_{b_{2g}}          \\
   -z F^{(1)}_{b_{1u}}-z(z^2\!-\!\frac{3}{5}) F^{(3)}_{b_{1u}}+x y F^{(2)}_{b_{1g}}   & F^{(0)}_{a_g^{yy}}+\!(y^2\!-\!\frac{1}{3})F^{(2)}_{a_g^{yyA}}+\!(z^2\!-\!x^2)F^{(2)}_{a_g^{yyB}}  & x F^{(1)}_{b_{3u}}+x(x^2\!-\!\frac{3}{5}) F^{(3)}_{b_{3u}}+y z F^{(2)}_{b_{3g}}          \\
    y F^{(1)}_{b_{2u}}+y(y^2\!-\!\frac{3}{5}) F^{(3)}_{b_{2u}}+x z F^{(2)}_{b_{2g}}   &-x F^{(1)}_{b_{3u}}-x(x^2\!-\!\frac{3}{5}) F^{(3)}_{b_{3u}}+y z F^{(2)}_{b_{3g}}                   & F^{(0)}_{a_g^{zz}}+\!(z^2\!-\!\frac{1}{3})F^{(2)}_{a_g^{zzA}}+\!(x^2\!-\!y^2)F^{(2)}_{a_g^{zzB}} \\
  \end{array}
\right)
\end{eqnarray}
\end{widetext}

\section{Polarization dependence of scattered intensity}

The scattering tensor can not be measured directly. Absorption, reflection, scattering or diffraction experiments measure different parts or combinations of the tensor. An absorption measurement probes the imaginary part of $\sigma$.
\begin{equation}
I_{abs}=-\Im\left[\sum_i \bm{\varepsilon}\cdot\sigma_i\cdot\bm{\varepsilon}\right]
\end{equation}
with $\bm{\varepsilon}$ the polarization of the light, $\sigma_i$ the conductivity tensor of atom $i$. The sum is over all atoms in the sample (neglecting self absorption effects) and $\sigma_i\propto F_i/\omega$. The real part can then be obtained from a Kramers-Kronig transformation.

In a reflection, scattering or diffraction experiment one measures
\begin{equation}
I_{scat}=\left| \sum_i e^{\imath (\bm{k_{in}}-\bm{k_{out}})\cdot\bm{r}_i} \bm{\varepsilon}_{out}\cdot F_i\cdot\bm{\varepsilon}_{in} \right|^2
\end{equation}
with $\bm{\varepsilon}_{in(out)}$ the polarization of the incoming (outgoing) light, $\bm{k}_{in(out)}$ the wave vector of the incoming (outgoing) light, $\bm{r}_i$ the position of atom $i$, $F_i$ the scattering tensor of atom $i$ and the sum over all atoms in the sample (neglecting self absorption effects).

The polarization can not be chosen arbitrarily in an diffraction experiment, as the $\bm{k}$-vectors of the light must fulfill the Bragg condition. The polarization has to be perpendicular to $\bm{k}$, leaving two options for the polarization: $\bm{\varepsilon}$ can be in the scattering plane ($\bm{\pi}$ polarization) or $\bm{\varepsilon}$ can be perpendicular to the scattering plane ($\bm{\sigma}$ polarization). Theoretically it is easier to work in Cartesian coordinates and describe the material properties independent of the measurement geometry. We express the magnetization direction, the Pointing vectors of the light as well as the polarization in the same coordinate system. It is thus needed to find an easy expression of $\bm{\sigma}$ and $\bm{\pi}$ polarization in this coordinate frame. We will take that $\bm{a}$, $\bm{b}$, and $\bm{c}$, (the lattice vectors) and the scattering vector ($\bm{q}\equiv[q_x,q_y,q_z]$) are known in real space Cartesian coordinates.\cite{vector} In order to define the scattering plane we will need a second vector that might arbitrarily be chosen, but must be perpendicular to $\bm{q}$, which we will call $\bm{q}_{\perp}$. Then $\phi$ will be the azimuthal angle, defined as the angle between $\bm{q}_{\perp}$ and the scattering plane. The angle between $\bm{q}$ and $\bm{k}_{in}$ will be written as $\theta'$. (The angle between $\bm{q}$ and $\bm{k}_{out}$ then is $\pi-\theta'$). This leads to the following definitions for the $\bm{k}$ vectors of the light and $\bm{\sigma}$ and $\bm{\pi}$ polarization in Cartesian coordinates.
\begin{eqnarray}
\nonumber \bm{k}_{in}   &\parallel&\sin(\theta')(\cos(\phi)\bm{\hat{q}}_{\perp}+\sin(\phi)(\bm{\hat{q}}_{\perp}\times\bm{\hat{q}}))+\cos(\theta')\bm{\hat{q}}\\
\nonumber \bm{k}_{out}  &\parallel&\sin(\theta')(\cos(\phi)\bm{\hat{q}}_{\perp}+\sin(\phi)(\bm{\hat{q}}_{\perp}\times\bm{\hat{q}}))-\cos(\theta')\bm{\hat{q}}\\
\nonumber \bm{\sigma}   &    =    &(\bm{\hat{k}}_{in}\times\bm{\hat{k}}_{out})/\sin(2\theta')\\
\nonumber \bm{\pi}_{in} &    =    &(\bm{\hat{k}}_{in}\times\bm{\sigma})\\
          \bm{\pi}_{out}&    =    &(\bm{\hat{k}}_{out}\times\bm{\sigma})
\end{eqnarray}
The polarization of the incoming (outgoing) light can be written in Cartesian coordinates as $\bm{\varepsilon}_{in(out)}=\alpha \bm{\sigma} + \beta \bm{\pi}_{in(out)}$, with $\alpha$ and $\beta$ complex numbers such that $\alpha^*\alpha+\beta^*\beta=1$.

The intensity can be obtained in terms of dot and cross products of the polarization and a unit vector in the local moment direction ($\bm{\hat{m}}$). Hannon and Blume \textit{et al.} \cite{Hannon88} came to the following formula in spherical symmetry.
\begin{eqnarray}
F_{\bm{\varepsilon}_{in}\bm{\varepsilon}_{out}}&=&F^{(0)}(\bm{\varepsilon}_{in}\cdot\bm{\varepsilon}_{out}^*)\\
\nonumber                                      &+&F^{(1)}(\bm{\varepsilon}_{in}\times\bm{\varepsilon}_{out}^*\cdot\bm{\hat{m}})\\
\nonumber                                      &+&F^{(2)}((\bm{\varepsilon}_{out}^*\cdot\bm{\hat{m}})(\bm{\varepsilon}_{in}\cdot\bm{\hat{m}})-\frac{1}{3}(\bm{\varepsilon}_{in}\cdot\bm{\varepsilon}_{out}^*))
\end{eqnarray}
which can be derived by dotting the scattering tensor in spherical symmetry for arbitrary spin direction with $\bm{\varepsilon}_{out}^*$ and $\bm{\varepsilon}_{in}$ and some algebra.

In cubic ($O_h$) symmetry $F_{\bm{\varepsilon}_{in}\bm{\varepsilon}_{out}}$ becomes
\begin{eqnarray}
                                               &=&F^{(0)}_{a_{1g}}(\bm{\varepsilon}_{in}\cdot\bm{\varepsilon}_{out}^*)\\
\nonumber                                      &+&F^{(1)}_{t_{1u}}(\bm{\varepsilon}_{in}\times\bm{\varepsilon}_{out}^*\cdot\bm{\hat{m}})\\
\nonumber                                      &+&F^{(2)}_{t_{2g}}((\bm{\varepsilon}_{out}^*\cdot\bm{\hat{m}})(\bm{\varepsilon}_{in}\cdot\bm{\hat{m}})-(\bm{\varepsilon}_{out}^**\bm{\hat{m}})\cdot(\bm{\varepsilon}_{in}*\bm{\hat{m}}))\\
\nonumber                                      &+&F^{(2)}_{e_g}((\bm{\varepsilon}_{out}^**\bm{\hat{m}})\cdot(\bm{\varepsilon}_{in}*\bm{\hat{m}})-\frac{1}{3}(\bm{\varepsilon}_{in}\cdot\bm{\varepsilon}_{out}^*))\\
\nonumber                                      &+&F^{(3)}_{t_{1u}}(\bm{\varepsilon}_{in}\times\bm{\varepsilon}_{out}^*\cdot(\bm{\hat{m}}*(\bm{\hat{m}}*\bm{\hat{m}}-\frac{3}{5})))
\end{eqnarray}
with "$\times$" the standard vector cross product, "$\cdot$" the vector dot product and "$*$" stands for vector multiplication per index ($\bm{a}=\bm{b}*\bm{c} \Leftrightarrow a_i=b_i c_i \forall i$).

In tetragonal symmetry ($D_{4h}$) one obtains
\begin{eqnarray}
                                               &=&F^{(0)}_{a_{1g}^B}(\bm{\varepsilon}_{in_{xy}}\cdot\bm{\varepsilon}_{out_{xy}}^*)\\
\nonumber                                      &+&F^{(0)}_{a_{1g}^A}(\bm{\varepsilon}_{in_z}\cdot\bm{\varepsilon}_{out_z}^*)\\
\nonumber                                      &+&F^{(1)}_{e_u}(\bm{\varepsilon}_{in}\times\bm{\varepsilon}_{out}^*\cdot\bm{\hat{m}}_{xy})\\
\nonumber                                      &+&F^{(1)}_{a_{2u}}(\bm{\varepsilon}_{in}\times\bm{\varepsilon}_{out}^*\cdot\bm{\hat{m}}_{z})\\
\nonumber                                      &+&F^{(2)}_{e_g}((\bm{\varepsilon}_{out}^*\cdot\bm{\hat{m}}_{xy})(\bm{\varepsilon}_{in}\cdot\bm{\hat{m}}_{z})+(\bm{\varepsilon}_{out}^*\cdot\bm{\hat{m}}_{z})(\bm{\varepsilon}_{in}\cdot\bm{\hat{m}}_{xy}))\\
\nonumber                                      &+&F^{(2)}_{b_{2g}}((\bm{\varepsilon}_{out}^*\cdot\bm{\hat{m}}_{x})(\bm{\varepsilon}_{in}\cdot\bm{\hat{m}}_{y})+(\bm{\varepsilon}_{out}^*\cdot\bm{\hat{m}}_{y})(\bm{\varepsilon}_{in}\cdot\bm{\hat{m}}_{x}))\\
\nonumber                                      &+&F^{(2)}_{b_{1g}}(\frac{1}{2}(m_x^2-m_y^2)(\bm{\varepsilon}_{in_x}\cdot\bm{\varepsilon}_{out_x}^*-\bm{\varepsilon}_{in_y}\cdot\bm{\varepsilon}_{out_y}^*)\\
\nonumber                                      &+&F^{(2)}_{a_{1g}^A}(m_z^2-\frac{1}{3})(\bm{\varepsilon}_{in_z}\cdot\bm{\varepsilon}_{out_z}^*)\\
\nonumber                                      &+&F^{(2)}_{a_{1g}^B}(m_z^2-\frac{1}{3})(\bm{\varepsilon}_{in_{xy}}\cdot\bm{\varepsilon}_{out_{xy}}^*)\\
\nonumber                                      &+&F^{(3)}_{e_{u}}(\bm{\varepsilon}_{in}\times\bm{\varepsilon}_{out}^*\cdot(\bm{\hat{m}}_{xy}*(\bm{\hat{m}}*\bm{\hat{m}}-\frac{3}{5})))\\
\nonumber                                      &+&F^{(3)}_{a_{2u}}(\bm{\varepsilon}_{in}\times\bm{\varepsilon}_{out}^*\cdot(\bm{\hat{m}}_{z}*(\bm{\hat{m}}*\bm{\hat{m}}-\frac{3}{5})))
\end{eqnarray}
with $\bm{\hat{m}}_z$ a vector with only the $z$ component of $\bm{\hat{m}}$ non-zero, i.e. $[0,0,m_z/|\bm{m}|]$.
The orthorhombic ($D_{2h}$) relations are
\begin{eqnarray}
                                               &=&F^{(0)}_{a_g^{xx}}(\bm{\varepsilon}_{in_x}\cdot\bm{\varepsilon}_{out_x}^*)\\
\nonumber                                      &+&F^{(0)}_{a_g^{yy}}(\bm{\varepsilon}_{in_y}\cdot\bm{\varepsilon}_{out_y}^*)\\
\nonumber                                      &+&F^{(0)}_{a_g^{zz}}(\bm{\varepsilon}_{in_z}\cdot\bm{\varepsilon}_{out_z}^*)\\
\nonumber                                      &+&F^{(1)}_{b_{1u}}(\bm{\varepsilon}_{in}\times\bm{\varepsilon}_{out}^*\cdot\bm{\hat{m}}_{z})\\
\nonumber                                      &+&F^{(1)}_{b_{2u}}(\bm{\varepsilon}_{in}\times\bm{\varepsilon}_{out}^*\cdot\bm{\hat{m}}_{y})\\
\nonumber                                      &+&F^{(1)}_{b_{3u}}(\bm{\varepsilon}_{in}\times\bm{\varepsilon}_{out}^*\cdot\bm{\hat{m}}_{x})\\
\nonumber                                      &+&F^{(2)}_{b_{1g}}((\bm{\varepsilon}_{out}^*\cdot\bm{\hat{m}}_{x})(\bm{\varepsilon}_{in}\cdot\bm{\hat{m}}_{y})+(\bm{\varepsilon}_{out}^*\cdot\bm{\hat{m}}_{y})(\bm{\varepsilon}_{in}\cdot\bm{\hat{m}}_{x}))\\
\nonumber                                      &+&F^{(2)}_{b_{2g}}((\bm{\varepsilon}_{out}^*\cdot\bm{\hat{m}}_{z})(\bm{\varepsilon}_{in}\cdot\bm{\hat{m}}_{x})+(\bm{\varepsilon}_{out}^*\cdot\bm{\hat{m}}_{x})(\bm{\varepsilon}_{in}\cdot\bm{\hat{m}}_{z}))\\
\nonumber                                      &+&F^{(2)}_{b_{3g}}((\bm{\varepsilon}_{out}^*\cdot\bm{\hat{m}}_{y})(\bm{\varepsilon}_{in}\cdot\bm{\hat{m}}_{z})+(\bm{\varepsilon}_{out}^*\cdot\bm{\hat{m}}_{z})(\bm{\varepsilon}_{in}\cdot\bm{\hat{m}}_{y}))\\
\nonumber                                      &+&(F^{(2)}_{a_g^{xxA}}(m_z^2-\frac{1}{3})+F^{(2)}_{a_g^{xxB}}(m_x^2-m_y^2))(\bm{\varepsilon}_{in_x}\cdot\bm{\varepsilon}_{out_x}^*)\\
\nonumber                                      &+&(F^{(2)}_{a_g^{yyA}}(m_z^2-\frac{1}{3})+F^{(2)}_{a_g^{yyB}}(m_x^2-m_y^2))(\bm{\varepsilon}_{in_y}\cdot\bm{\varepsilon}_{out_y}^*)\\
\nonumber                                      &+&(F^{(2)}_{a_g^{zzA}}(m_z^2-\frac{1}{3})+F^{(2)}_{a_g^{zzB}}(m_x^2-m_y^2))(\bm{\varepsilon}_{in_z}\cdot\bm{\varepsilon}_{out_z}^*)\\
\nonumber                                      &+&F^{(3)}_{b_{1u}}(\bm{\varepsilon}_{in}\times\bm{\varepsilon}_{out}^*\cdot(\bm{\hat{m}}_z*(\bm{\hat{m}}*\bm{\hat{m}}-\frac{3}{5})))\\
\nonumber                                      &+&F^{(3)}_{b_{2u}}(\bm{\varepsilon}_{in}\times\bm{\varepsilon}_{out}^*\cdot(\bm{\hat{m}}_y*(\bm{\hat{m}}*\bm{\hat{m}}-\frac{3}{5})))\\
\nonumber                                      &+&F^{(3)}_{b_{3u}}(\bm{\varepsilon}_{in}\times\bm{\varepsilon}_{out}^*\cdot(\bm{\hat{m}}_x*(\bm{\hat{m}}*\bm{\hat{m}}-\frac{3}{5})))
\end{eqnarray}

 \begin{figure*}[h!t]
    \includegraphics[width=0.83\textwidth]{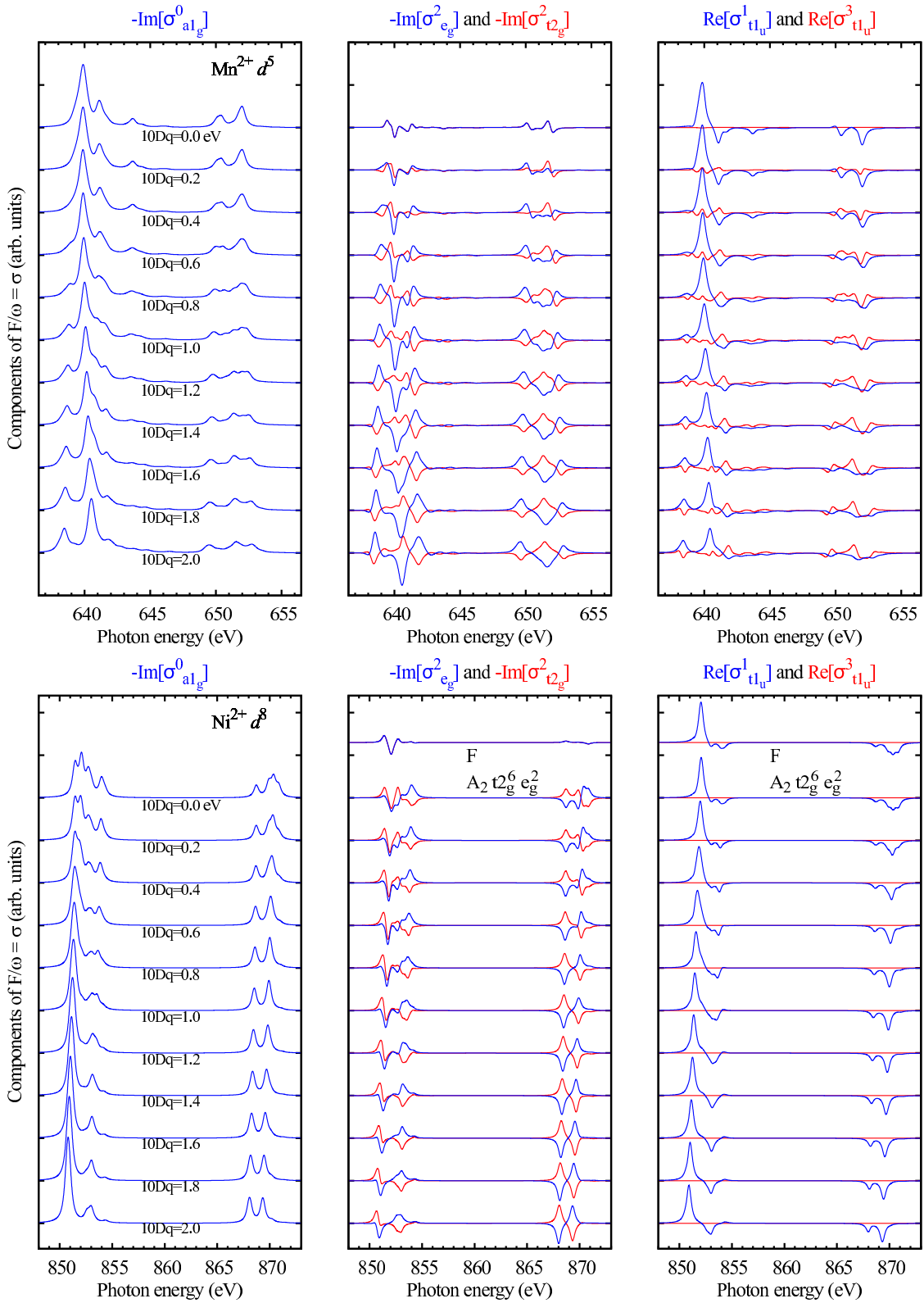}
    \caption{(color online) Scattering tensor of (top) Mn$^{2+}$ and (bottom) Ni$^{2+}$ as a function of the cubic crystal field splitting $10Dq$. Left column shows the $a1_{g}$ component, middle column the $t_{2g}$ and $e_g$ component and right column the $t_{1u}$ component. All graphs are on the same intensity scale. For Ni$^{2+}$ with 10Dq=0 two graphs are included, once starting from a spherical ground-state (F) and once starting from the cubic ground state (A2). Calculations done on an ionic model without spin-orbit coupling on the $d$-shell. \newline}
    \label{OhMnNi}
  \end{figure*}

\section{Mn$^{2+}$ and Ni$^{2+}$ in cubic symmetry}

The question that still has to be answered is whether the differences in intensity between spherical and lower symmetries are large enough to be observed. We therefore calculated a few examples at the transition metal $L_{2,3}$ edge using the multiplet crystal field approach. The parameters for such calculations are well discussed in the literature, \cite{Tanaka94} and we adopted those values. In order to show the evolution from spherical to cubic and from cubic to tetragonal symmetry we calculated the spectra for several crystal-field parameters. We did not include the $3d$ spin-orbit coupling nor covalency (crystal-field theory instead of ligand-field theory) in order to make the interpretation easier. While we did not include spin orbit coupling on the $d$ shell the magnetization direction directly coincides with the spin direction. The spectra are calculated at 0 K and assume a full magnetization. The calculations have been done with the full multiplet ligand field theory program XTLS 8.3. \cite{Tanaka94} The fundamental spectra can be obtained by doing calculations for specific spin orientations. On top of those calculations a basic check has been made where the optical conductivity tensor has been calculated for many different magnetization directions. The result is expanded on spherical harmonics in coordinates of the magnetization direction. Both methods gave the same results within the numerical accuracy of the calculations.

In the top panel of Fig. \ref{OhMnNi} we show the $L_{2,3}$ edge x-ray absorption spectra of Mn$^{2+}$ as a function of the cubic splitting parametrized by $10Dq$. The top spectrum has $10Dq=0$ and is therefore in spherical symmetry. The ground-state has always 5 electrons with parallel spin occupied and the ground-state wave function is independent of the crystal-field splitting. All changes in the spectra are due to final-state effects, whereby the $2p$ core hole is excited into a $t_{2g}$ or $e_g$ electron and the energy difference between the two changes as a function of $10Dq$. The left panel shows $F^{(0)}_{a_{1g}}$, which is the isotropic spectrum and the spectrum of the para-magnetic phase. These spectra are the same as published earlier by De Groot \textit{et al.} \cite{deGroot90}. The middle column shows $F^{(2)}_{e_g}$ and $F^{(2)}_{t_{2g}}$. These two spectra describe the magnetic linear dichroic effect. In spherical symmetry (top curve, $10Dq=0$) these two spectra are the same. However when a cubic distortion is present these two spectra become different. For a commensurate spin spiral one would expect a different resonant profile at the second order Bragg peak depending on the phase of the spiral with respect to the cubic crystal structure. i.e. a spiral with period $4z$ and the spins in the $x$, $y$, $-x$, $-y$ direction would show an $|F^{(2)}_{e_g}|^2$ like second order resonance profile, whereas a spiral at the same $q$ vector, but shifted in phase such that the spin directions are $x+y$, $-x+y$, $-x-y$, $x-y$ would show an $|F^{(2)}_{t_{2g}}|^2$ like second order resonance profile. For x-ray absorption this has implications for the measured magnetic linear dichroism as explained previous by Arenholz and Van der Laan \textit{et al.}\cite{Arenholz06, Arenholz07, Laan08}. The right column shows the $F^{(1)}_{t_{1u}}$ and $F^{(3)}_{t_{1u}}$ spectra. In $O_h$ symmetry the $k=1$ spectra branch to a single representation (a $t_{1u}$). Additionally angular momentum ($k$) is not a good quantum number in cubic symmetry and therefore the $F^{(3)}_{t_{1u}}$ spectra becomes non-zero. For spectroscopy this means a difference in the magnetic circular dichroism spectra for a system magnetized in the [001] direction where one measures $\Re[\sigma_{t_{1u}}^{1}+2/5\sigma_{t_{1u}}^{3}]$ and the [111] direction where the measured spectrum is $\Re[\sigma_{t_{1u}}^{1}-4/15\sigma_{t_{1u}}^{3}]$. For resonant diffraction at an anti-ferromagnetic Bragg reflection this will lead to a different resonance profile depending if the spins are ordered in the [001] or [111] direction. It will also lead to a different azimuthal dependence of the scattered intensity at an anti-ferromagnetic Bragg reflection. For spins oriented in the [$11\bar{2}$] direction ferromagnetically aligned in (111) planes, antiferromagnetically stacked in the [$111$] direction and a scattering geometry such that $\theta'=\pi/6$ one finds a scattered intensity proportional to $|F^{(3)}/(2\sqrt6)\pm\sin(\phi)(10F^{(1)}-F^{(3)})/20|^2$ in the $\pi$-$\sigma$ ($\sigma$-$\pi$) scattering channel. For different spin alignments the $\phi$ independent scattering due to the $F^{(1)}$ scattering tensor would be related to the projection of the spin on the $\bm{q}$ vector. Measurements at a single resonant energy could thus lead to incorrect determination of the spin direction. However $F^{(3)}$ has a different $\omega$ dependence than $F^{(1)}$ and this can be used to obtain the correct spin direction from such experiments.

The bottom panel of Fig. \ref{OhMnNi} shows the same spectra, for Ni$^{2+}$. The major difference between a $d^8$ system (Ni$^{2+}$) and a $d^5$ system (Mn$^{2+}$) is that for a $d^8$ system the ground-state is modified when a cubic crystal-field is included. In spherical symmetry both the $e_g$ and $t_{2g}$ orbitals are equally occupied, in cubic symmetry this is not the case. When $10Dq=0$ and a spherical charge distribution is assumed for the initial state one finds that the $F^{(2)}_{e_g}$ and $F^{(2)}_{t_{2g}}$ spectra are equivalent (middle panel top curve labeled as $F$). If one however assumes a $t_{2g}^6 e_g^2$ orbital occupation as one would find as a ground-state in cubic symmetry one finds a large difference between the $F^{(2)}_{e_g}$ and $F^{(2)}_{t_{2g}}$ spectra (middle panel second curve from top labeled as $A_2$). It is important to note the difference between Mn$^{2+}$ and Ni$^{2+}$. For a perturbation that does not break the ground-state symmetry the changes in spectra scale with the size of the distortion, whereas for a perturbation that breaks the ground-state symmetry there will be an immediate large effect on the spectral line-shape as long as the distortion is bigger than temperature. The symmetry breaking is important for orbitally ordered systems. Above the ordering temperature, the system can be described by the scattering tensor in the higher symmetry. Although the distortions may be small in the orbitally ordered phase, breaking the ground state symmetry will have an immediate effect on the spectra. Thus even small distortions lead to relevant changes in the scattering tensor.

\section{Ni$^{2+}$, Mn$^{3+}$ and Cu$^{2+}$ in tetragonal symmetry}

\begin{figure*}[h!t]
    \includegraphics[width=0.85\textwidth]{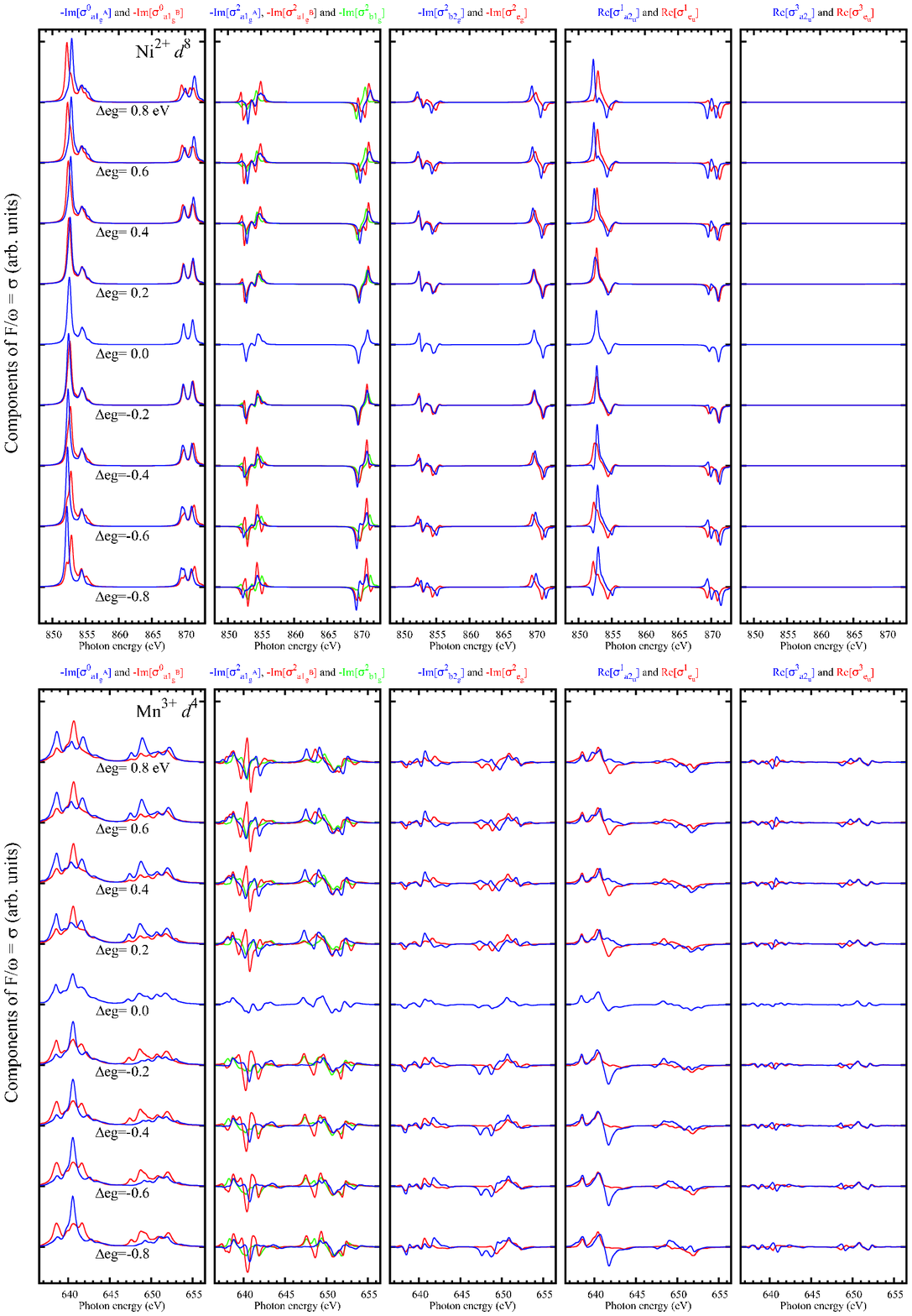}
    \caption{continues on next page \newline \newline}
\end{figure*}
\begin{figure*}[h!t]
    \ContinuedFloat
    \includegraphics[width=0.85\textwidth]{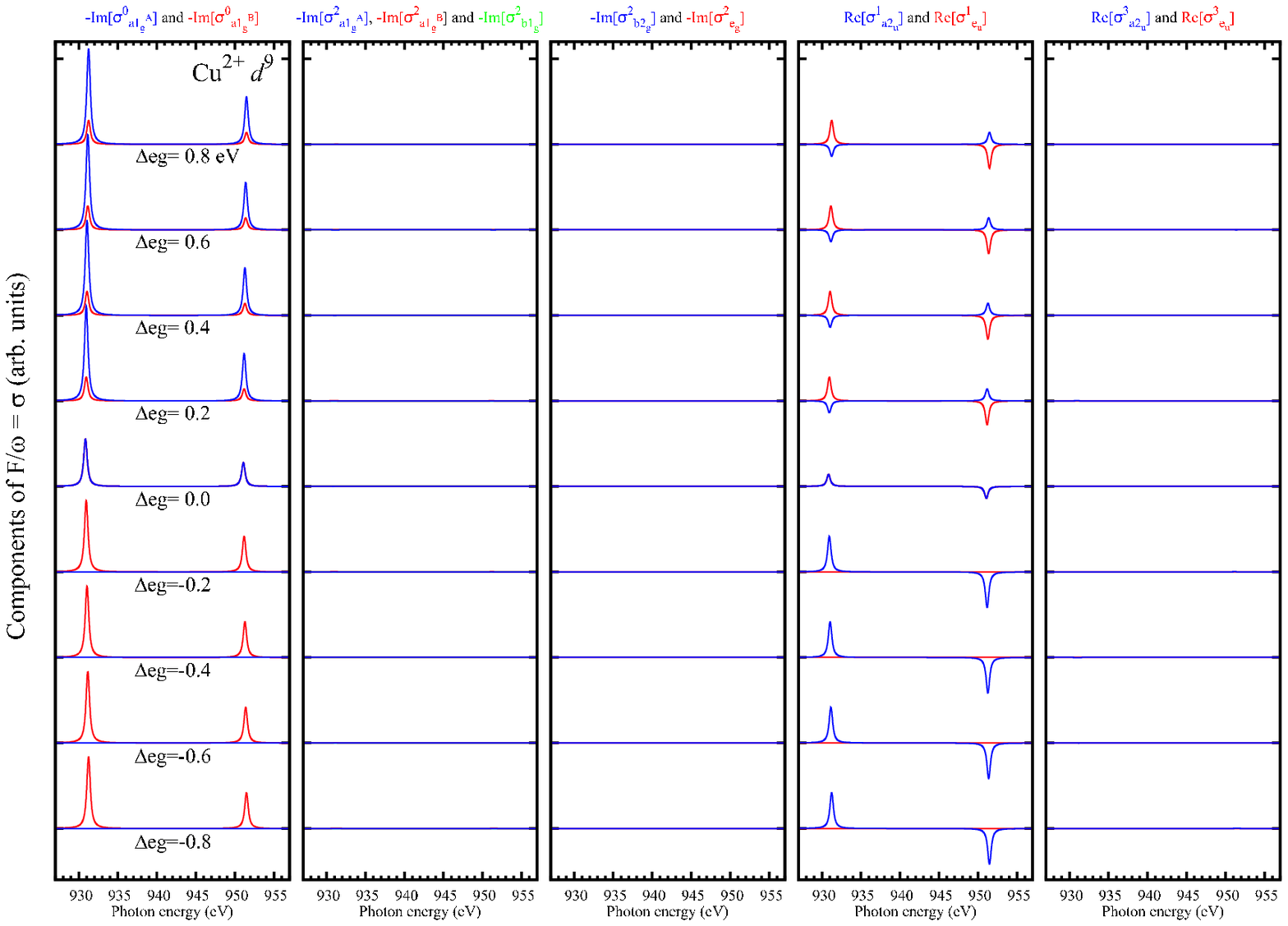}
    \caption{(color online) Fundamental spectra of the elastic scattering tensor for Ni$^{2+}$ (top panel), Mn$^{3+}$ (middle panel) and Cu$^{3+}$ (bottom panel) as a function of tetragonal crystal-field splitting $\Delta$e$_g$ defined as the energy difference between the $d_{z^2}$ and $d_{x^2-y^2}$ orbital. ($\Delta$t$_{2g}$=1/4$\Delta$e$_g$). $\Delta$eg$>0$ stands for a tetragonal contracted system, i.e. the $d_{z^2}$ orbital higher in energy than the $d_{x^2-y^2}$ orbital.}
    \label{D4hNiMnCu}
\end{figure*}

In tetragonal symmetry the $F^{(0)}$ spectrum branches to $F^{(0)}_{a_{1g}^A}$ and $F^{(0)}_{a_{1g}^B}$, the $F^{(1)}$ spectrum branches to $F^{(1)}_{a_{2u}}$ and $F^{(1)}_{a_{eu}}$, and the $F^{(2)}$ spectrum branches to $F^{(2)}_{b_{2g}}$, $F^{(2)}_{e_{g}}$, $F^{(2)}_{b_{1g}}$, $F^{(2)}_{a_{1g}^A}$ and $F^{(2)}_{a_{1g}^B}$. To estimate the numerical differences between these spectra at the $L_{2,3}$ edge we again did crystal field calculations.

The top panel of Fig. \ref{D4hNiMnCu} shows the spectra for Ni$^{2+}$ as a function of tetragonal distortion. In the middle we show cubic symmetry and towards the top (bottom) a tetragonal contraction (elongation). The Ni$^{2+}$ ground-state does not change as a function of tetragonal distortion and therefore the difference between spectra that branch from the same irreducible representation in cubic symmetry scale with the size of the tetragonal distortion. The left panel shows the natural linear dichroism in tetragonal distorted Ni$^{2+}$ which has been discussed before. \cite{haverkort04} The second panel from the left shows the linear dichroism that branches from the cubic $e_g$ representation. Interestingly, \emph{three} fundamental spectra represent the linear dichroism that branches from the $e_g$ representation in cubic symmetry and for feasible distortions they are all different. The trace of the conductivity tensor, or the isotropic spectrum, is thus dependent on the magnetization direction and the difference is large enough to be detectable. The isotropic spectrum in $D_{4h}$ symmetry is given by $(1/3)(2F_{a_{1g}^B}^{(0)}+F_{a_{1g}^A}^{(0)}+(z^2-1/3)(F_{a_{1g}^A}^{(2)}-F_{a_{1g}^B}^{(2)})$ and $(F_{a_{1g}^A}^{(2)}-F_{a_{1g}^B}^{(2)})$ is, as clearly can be seen in Fig. \ref{D4hNiMnCu}, non zero. The linear dichroism that branches from the cubic $t_{2g}$ representation is shown in the middle panel. The second panel from the right shows the circular dichroic spectra. Differences in the antisymmetric part of the scattering tensor are almost always neglected, but for reasonable distortions, present in e.g. layered perovskite structures large differences can be seen between the $F^{(1)}_{b_{2u}}$ and $F^{(1)}_{e_u}$ spectra. Going beyond $O_{h}$ symmetry thus also affects the antisymmetric part of $F$.

The middle panel of Fig. \ref{D4hNiMnCu} shows the spectra of Mn$^{3+}$ as a function of tetragonal distortion. Mn$^{3+}$ has a $d^4$ configuration and depending on the distortion it will have an occupied $d_{x^2-y^2}$ orbital (top) or an occupied $d_{z^2}$ orbital. As in the case of Ni$^{2+}$, for the spherical to cubic distortion, one has an immediate effect for all fundamental spectra and in particular also in the antisymmetric part of the scattering tensor. This is important for magnetic circular dichroism and RXD in all kinds of manganates where orbital order and magnetic order co-exists.

The bottom panel of Fig. \ref{D4hNiMnCu} shows the fundamental spectra of the scattering tensor for Cu$^{2+}$ a $d^9$ configuration. This is a particularly simple and instructive example as all calculations can be done by hand and everything can be done in a one particle picture. The ground-state has one hole in the $d$ shell and the final state one hole in the $2p$ shell. The ground-state hole in the $d$-shell will either be in the $x^2-y^2$ orbital (bottom of Fig. \ref{D4hNiMnCu}) or in the $z^2$  orbital (top of Fig. \ref{D4hNiMnCu}). For the natural dichroism (left panel, $F^{(0)}_{a_{1g}^{xx}}$ and $F^{(0)}_{a_{1g}^{zz}}$) one finds that for an $d_{x^2-y^2}$ hole there is no intensity in the $F^{(0)}_{a_{1g}^{zz}}$ as one can not excite into the $d_{x^2-y^2}$ electron with $z$ polarized light. For an $d_{z^2}$ hole there is a ratio of 1 to 3 for the different fundamental spectra. Note that the spectra are independent of the size of the distortion and only depend on the symmetry of the ground-state. The symmetric part of the magnetic scattering tensor is zero for a $d^9$ configuration, which is a general property of any Kramers doublet. The spin-up and spin-down state are related by a mirror symmetry and therefore should have the same symmetric part of the scattering tensor. At the same time the sum of the scattering tensor for the spin-up and spin-down state should be equal to the paramagnetic tensor which is given by the $F^{(0)}$ components. Therefore the symmetric magnetic part of the scattering tensor should be zero for a ground-state Kramers doublet. The antisymmetric part shows a very strong crystal orientation dependence. For a $d_{x^2-y^2}$ hole in the ground-state, the $F^{(1)}_{b_{2u}}$ spectra is finite, but the $F^{(1)}_{e_u}$ spectrum is zero. This is again related to the fact that one can not excite into the $d_{x^2-y^2}$ hole with $z$ polarized light. For the cuprates, the RXD spectra are thus only sensitive to the out-of-plane moment. 

\section{Conclusion}

We have shown that the fundamental spectra of the scattering or conductivity tensor are strongly influenced by the crystal field symmetry at the transition metal $L_{2,3}$ edge. Both the symmetric part as noted previously, \cite{Arenholz06, Arenholz07,Laan08} but also the antisymmetric part show strong deviations from the spectra calculated in spherical symmetry. When the symmetry is lowered the fundamental spectra known from spherical symmetry branch into several different spectra. At the same time higher order spectra become non-zero and are important. In general more fundamental spectra are needed in order to describe the conductivity tensor as one might have expected. We have shown to what extent the crystal symmetry effects the magneto-optical effects and how they influence RXD and XAS experiments.

The reason for the deviations from the simple rule that the scattered intensity at an antiferromagnetic Bragg reflection is proportional to $\left|(\bm{\varepsilon}_{in}\times\bm{\varepsilon}_{out}^*)\cdot\bm{\hat{m}}\right|^2$ is given by the fact that in spherical symmetry there is only one tensor element that describes the antisymmetric part of the conductivity tensor. In cubic symmetry this already becomes more complicated as $F^{(3)}$ can not be neglected if $S$ is larger then 1. In tetragonal or lower symmetry also $F^{(1)}$ branches to different spectra. When the spin aligns in a high symmetry direction of the crystal there still will be only one tensor element describing the antisymmetric part of $F$ and therefore the simple scattering rules as derived in spherical symmetry will still hold. Even when this tensor element is made from a linear combination of $F^{(1)}$ and $F^{(3)}$; as long as their ratio is the same for all polarizations the simple rules hold. In most cases the, single ion anisotropy will dictate the preferred direction of the magnetic moments, aligning it along high symmetry directions. In this case the simple rules are valid. 

Local magnetic moments are not always aligned in a high-symmetry crystal direction. Well known examples are the simple rock-salts like NiO, MnO and CoO. An other class would be magnetic spiral structures as found in the cuprates and manganates. In these cases a more involved spectral line shape analysis is needed in order to determine the magnetization direction. One then can not measure the scattered intensity at a single photon energy, but needs to measure the energy dependence at all azimuthal directions in order to obtain more information. By comparing the line-shape to theory at different azimuthal directions or by the use of sum-rules one then can still obtain the full information as present in the data.

The present calculations could be proved by magnetic circular and linear dichroism experiments. Temperature dependence of the magneto-optical effects as well as non-full magnetization of the sample will be discussed in a future work.

\section{Acknowledgments}

The research in K{\"o}ln is supported by the Deutsche Forschungsgemeinschaft through Grant No. SFB 608 and the Cologne Bonn Graduate School.


\begin{thebibliography}{99}
\bibitem{vector} Within this paper the notation $[x,y,z]$ is used to define a vector in real space Cartesian coordinates. The vector $[x,y,z]$ is thus not necessarily equal to the vector $x \bm{a}$+$y \bm{b}$+$z \bm{c}$ with $\bm{a}$, $\bm{b}$, and $\bm{c}$ the primitive vectors spanning the Bravis lattice.
\end{thebibliography}
\end{document}